\documentclass[aps,prb,reprint]{revtex4-1}
\usepackage{amsmath,amssymb}
\usepackage{graphicx}
\usepackage[caption=false]{subfig}

\usepackage[uline]{hhtensor}

\newcommand\ed{\mathbf{p}}
\newcommand\md{\mathbf{m}}
\newcommand\rrp{\mathbf{r^{\prime}}}
\newcommand\rr{\mathbf{r}}

\newcommand\Erw{\mathbf{E}(\rrp,\omega)}
\newcommand\Hrw{\mathbf{H}(\rrp,\omega)}

\newcommand\MdE{\matr{\alpha}_{\ed\mathbf{E}}}
\newcommand\MdH{\matr{\alpha}_{\ed\mathbf{H}}}
\newcommand\MmE{\matr{\alpha}_{\md\mathbf{E}}}
\newcommand\MmH{\matr{\alpha}_{\md\mathbf{H}}}

\newcommand\Eed{\mathbf{E_{\ed}}}
\newcommand\Emd{\mathbf{E_{\md}}}
\newcommand\Hed{\mathbf{H_{\ed}}}
\newcommand\Hmd{\mathbf{H_{\md}}}

\newcommand\edbar{\mathbf{\bar{p}}}
\newcommand\Eedbar{\mathbf{E_{\edbar}}}
\newcommand\Hedbar{\mathbf{H_{\edbar}}}

\newcommand\gmunu{g_{\mu\nu}}

\newcommand\E{\mathbf{E}}
\newcommand\HH{\mathbf{H}}

\newcommand\F{\mathcal{F}}
\newcommand\G{\mathcal{G}}

\newcommand\D{\mathbf{D}}
\newcommand\B{\mathbf{B}}

\begin{document}
\title{The role of duality symmetry in transformation optics}
\author{Ivan Fernandez-Corbaton$^{1,2}$ and Gabriel Molina-Terriza$^{1,2}$}
\affiliation{$^1$ Department of Physics \& Astronomy, Macquarie University, NSW 2109, Australia}
\affiliation{$^2$ ARC Center for Engineered Quantum Systems, NSW 2109, Australia}
\begin{abstract}
Maxwell's equations in curved space-time are invariant under electromagnetic duality transformations. We exploit this property to constrain the design parameters of metamaterials used for transformation electromagnetics. We show that a general transformation must be implemented using a dual-symmetric metamaterial. We also show that the spatial part of the coordinate transformation has the same action for both helicity components of the electromagnetic field, while the spatio-temporal part has an helicity dependent effect. Dual-symmetric metamaterials can be designed by constraining the polarisability tensors of their individual constituents, i.e. the meta atoms. We obtain explicit expressions for these constraints. Two families of realistically implementable dual symmetric meta atoms are discussed, one that exhibits electric-magnetic cross-polarisability and one that does not. In simple three dimensional periodical arrangements of the meta atoms (Bravais lattices), the helicity dependent effect can only be achieved if the meta atoms exhibit non-zero electric-magnetic cross-polarisabilities. In our derivations, we find that two dipoles located at the same point, one electric ($\ed$) and one magnetic ($\md$), are needed to produce a total field with well defined helicity equal to +1 or -1, and that they must be related as $\ed=\frac{i}{c}\md$ or $\ed=-\frac{i}{c}\md$, respectively.  
\end{abstract}

\maketitle
Transformation electromagnetics offers us a path to the design of invisibility cloaks, perfect lenses and any other device whose action on the electromagnetic field can be casted as a coordinate transformation in space-time \cite{Leonhardt2006,Pendry2006,Leonhardt2006b}. Transformation electromagnetics is based on the fact that Maxwell's equations in an empty region of curved space-time are equivalent to those inside a material medium in a flat space-time background \cite{Plebanski1960}. The desired transformation specifies a space-time metric which at its turn specifies the constitutive relations of the material. A detailed treatise in transformation electromagnetics can be found in \cite{Leonhardt2009}.

Such formidable step in our ability to manipulate electromagnetic waves comes with a correspondingly steep increase in the tunability requirements of material constitutive relations. Nature does not provide us with nearly enough flexibility in this aspect. We must synthesize artificial materials: Electromagnetic metamaterials \cite{Kock1956}. Transformation media are typically implemented by means of an ensemble of inclusions inside an homogeneous and isotropic dielectric. These inclusions, electromagnetically small for the wavelengths of the operating bandwidth, are sometimes referred to as meta atoms. The idea is to obtain the required constitutive relations from the collective response of the meta atoms. Currently, though, there is no systematic design methodology to go from the constitutive relations to the actual implementation of the metamaterial. In general, this is a highly complex task, partly because of the huge number of degrees of freedom of a general metamaterial, which include the electromagnetic response of the meta atoms and their three dimensional spatial arrangement. Reducing the number of degrees of freedom while maintaining the ability to implement general coordinate transformations is desirable.

 In this paper we use a non-geometrical symmetry of Maxwell's equations, electromagnetic duality, to constrain the individual response of the meta atoms without restricting the implementable transformations. There is a deep connection between transformation electromagnetics and duality symmetry. Almost two decades ago, I. Bialynicki-Birula realised that that the two helicity components of an electromagnetic wave do not mix in a gravitational field \cite{Birula1994,Birula1996}. Since helicity, as an operator, is the generator of duality transformations in the same sense that angular momentum is the generator of rotations, helicity preservation is equivalent to invariance under duality transformations. It follows that Maxwell's equations in a general space-time geometry are invariant under duality transformations. We can therefore obtain a constraint without sacrificing generality: a metamaterial that implements a transformation medium should preserve helicity, or in other words, possess duality invariance. It follows that any transformation can be implemented using only dual symmetric meta atoms, that is, meta atoms which, upon scattering, preserve the helicity of the electromagnetic field. Assuming that the meta atoms are electromagnetically small enough so that they can be treated in the dipolar approximation, we can obtain the restrictions that the requirement of duality invariance imposes on the polarisability tensors of the meta atoms. 

In this paper, we provide the theoretical basis and tools for the use of duality symmetry as a guide in the design of transformation electromagnetic devices. The paper starts with a brief introduction to the concepts of duality and helicity in free space and continues with the study of duality symmetry for materials with general linear constitutive relations. We obtain the restrictions that the constitutive relations must meet in order for the material to be dual symmetric. As expected, the constitutive relations induced by a general space time geometry meet those restrictions. Following the argument that the metamaterial must be dual symmetric, we derive the constrains on the polarisability tensor of a dual meta atom. Finally, we discuss two classes of realistically implementable dual symmetric meta atoms.

Electromagnetic duality is a transformation that mixes electric and magnetic fields by means of a real angle $\theta$ (Chap. 6.11 in Ref. \onlinecite{Jackson1998}). Assuming space and time dependent fields $(\rr,t)$: 
\begin{equation}
\label{eq:gendual}
\begin{split}
\mathbf{E}&\rightarrow \mathbf{E}_\theta=\mathbf{E}\cos\theta  - Z_0\mathbf{H}\sin\theta , \\
Z_0\mathbf{H}&\rightarrow Z_0\mathbf{H}_\theta=\mathbf{E}\sin\theta + Z_0\mathbf{H}\cos\theta ,
\end{split}
\end{equation}
where $Z_0=\sqrt{\mu_0/\epsilon_0}$ and $(e_0,\mu_0)$ are the vacuum permittivity and permeability constants. In vacuum, (\ref{eq:gendual}) is a symmetry of Maxwell's equations: If the electromagnetic field  $(\mathbf{E}(\rr,t),\mathbf{H}(\rr,t))$ is a solution of the free space Maxwell's equations, then the field $(\mathbf{E}_\theta(\rr,t),\mathbf{H}_\theta(\rr,t))$ is also a solution for any value of $\theta$. In the 1960's, Calkin \cite{Calkin1965} and Zwanziger \cite{Zwanziger1968} showed that helicity was the conserved quantity related to such symmetry. The helicity operator is defined (Chap. 8.4.1 in Ref. \onlinecite{Tung1985}) as the projection of the total angular momentum $\mathbf{J}$ onto the linear momentum direction, i.e. $\Lambda=\mathbf{J}\cdot\mathbf{P}/|\mathbf{P}|$. In the same way that linear momentum generates translations and angular momentum generates rotations, the helicity operator generates the duality transformation in (\ref{eq:gendual}). For the transverse electromagnetic field, helicity can take the values $\pm1$, which completely describe its polarisation degrees of freedom. It is possible to intuitively understand the meaning of helicity when considering the momentum space decomposition of a general field, that is, as a superposition of plane waves. In this representation, helicity is related to the handedness of the polarization of each and every plane wave. Helicity is well defined only when all the plane waves have the same handedness with respect to their momentum vector, including both propagating and evanescent plane waves. 

\begin{figure}[h]
\includegraphics[scale=0.45]{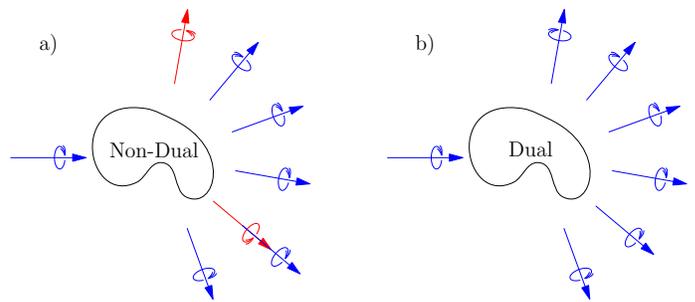}
\caption{(Color online) (a) In general, the helicity of an electromagnetic field is not preserved after interaction with a non-dual symmetric object. An incoming field with well defined helicity, in this case a single plane wave of definite polarisation handedness (blue), produces a scattered field that contains components of the opposite helicity (red). The helicity of the scattered field in figure (a) is not well defined because it contains plane waves of different helicities. (b) Helicity preservation after interaction with a dual symmetric object. The helicity of the scattered field is well defined an equal to the helicity of the input field.} 
\label{fig:dnd}
\end{figure}

In general, the presence of matter breaks the symmetry of the equations: a solution $(\mathbf{E}(\rr,t),\mathbf{H}(\rr,t))$ does not result in a new solution when transformed as in (\ref{eq:gendual}). As a consequence, the interaction with matter generally produces components of changed helicity. Nevertheless the electromagnetic duality symmetry can be restored for the source-free macroscopic Maxwell's equations in material systems characterized by scalar permittivities and permeabilities when these meet a particular constrain \cite{FerCor2012p}. 
We shall see that duality can also be restored in a macroscopic bianisotropic and inhomogeneous medium as well as in the dipolar approximation. Both of these cases are relevant for transformation devices made with metamaterials. Fig. \ref{fig:dnd} illustrates helicity preserving and non-preserving electromagnetic interactions with material systems.

In this paper, we will use the Riemann-Silberstein representation of electromagnetic fields \cite{Birula1994,Birula1996,Birula2013}. This formulism is very well suited to treat problems involving duality and helicity. If the duality transformation in (\ref{eq:gendual}) is applied to the combinations
\begin{equation}
\label{eq:g}
	\mathbf{G}_{\pm}(\rr,t)=\frac{1}{\sqrt{2}}\left(\E(\rr,t) \pm iZ_0\HH(\rr,t)\right),
\end{equation}
they transform in a simple way by just acquiring a phase $\mathbf{G}_{\pm}^\theta(\rr,t)\rightarrow \exp(\mp i\theta)\mathbf{G}_{\pm}(\rr,t)$. It means that the $\mathbf{G}_{\pm}(\rr,t)$ are eigenstates of the duality transformation with eigenvalues equal to $\exp(\mp i\theta)$. The same is true for 
\begin{equation}
\label{eq:f}
	\mathbf{F}_{\pm}(\rr,t)=\frac{1}{\sqrt{2}}\left(Z_0\D(\rr,t)\pm i\B(\rr,t)\right)
\end{equation}
under the companion transformation (Chap. 6.11 in Ref. \onlinecite{Jackson1998}):
{
\begin{equation}
\label{eq:gendual2}
\begin{split}
Z_0\mathbf{D}&\rightarrow Z_0\mathbf{D}_\theta=Z_0\mathbf{D}\cos\theta - \mathbf{B}\sin\theta, \\
\mathbf{B}&\rightarrow \mathbf{B}_\theta=Z_0\mathbf{D}\sin\theta + \mathbf{B}\cos\theta,
\end{split}
\end{equation}
}
which results in $\mathbf{F}_{\pm}^\theta(\rr,t)\rightarrow \exp(\mp i\theta)\mathbf{F}_{\pm}(\rr,t)$.

It follows that $\mathbf{G}_{\pm}(\rr,t)$ and $\mathbf{F}_{\pm}(\rr,t)$ are eigenstates of the generator of the duality transformation with eigenvalues equal to $\pm 1$. Under the restriction of positive frequencies, helicity can be shown to be such generator (see \cite{Calkin1965,Zwanziger1968,Deser1976,Cameron2012,FerCor2012p} for explicit derivations). As explained in \S 2.2 of \onlinecite{Birula1996}, this restriction avoids redundant degrees of freedom, and then $\mathbf{G}_+(\rr,t)$ and $\mathbf{F}_+(\rr,t)$ are eigenstates of the helicity operator with eigenvalue one and $\mathbf{G}_-(\rr,t)$ and $\mathbf{F}_-(\rr,t)$ are eigenstates of the helicity operator with eigenvalue minus one.

The crucial point is that the free-space time evolution of the two helicities $\mathbf{F}_{\pm}(\rr,t)$ or $\mathbf{G}_{\pm}(\rr,t)$ is decoupled \cite{Birula1996}. This is clearly seen in the free space Maxwell's curl equations written as \cite{Birula1996} ($c_0=1/\sqrt{\epsilon_0\mu_0}$):
\begin{equation}
\label{eq:curl}
i\partial_t \mathbf{F}_{\pm}(\rr,t)=\pm c_0 \nabla\times \mathbf{F}_{\pm}(\rr,t).
\end{equation}

The same arguments hold for the case of a homogeneous and isotropic medium with constitutive relations $\mathbf{D}=\epsilon\mathbf{E},\ \mathbf{B}=\mu\mathbf{H}$ by simply replacing  $\epsilon_0$ and $\mu_0$ by $\epsilon$ and $\mu$, hence $Z_0$ by $Z$ and $c_0$ by $c$.

The separate evolution of the two helicity components is a consequence of the invariance of the equations under duality transformations. This also happens for Maxwell's equations in a gravitational field. The formalism of transformation optics implicitly contains this invariance. This is the property that we exploit in the remainder of this paper to obtain guidelines in the design of metamaterials suitable for transformation devices.

Having used the very convenient Riemann-Silberstein notation to discuss helicity and duality in Maxwell's equations for electromagnetic fields of general time dependence, we will now assume a harmonic decomposition of all fields, as in
\begin{equation}
	\label{eq:X}
	\mathbf{X}(\mathbf{r},t)=\int_0^\infty d\omega \mathbb{R}\left\{\mathbf{\hat{X}}(\mathbf{r},\omega)\exp(-i\omega t)\right\},
\end{equation}
where $\mathbb{R}\{\cdot\}$ is the real part, and treat each frequency component $\mathbf{\hat{X}}(\mathbf{r},\omega)\exp(-i\omega t)$ separately for the rest of the paper. This separation is a common approach in the field of metamaterials because it allows for a simple treatment of the frequency dependent responses of many meta atoms. For linear systems, this setting is completely general. In (\ref{eq:X}), $\omega$ is restricted to positive frequencies only. It is important to note that this restriction does no reduce the generality of the treatment because, for electromagnetic fields, all the information is contained in the positive frequency part and duplicated in the negative frequency part, or vice versa. For the rest of the paper, the time dependence is always implicitly assumed to be harmonic. The general spatio-frequential dependency $(\rr,\omega)$ is assumed but normally not explicitly written. For example, the symbols $\mathbf{E}$ and $\mathbf{E}(\rr,\omega)$ both refer to $\mathbf{\hat{E}}(\mathbf{r},\omega)\exp(-i\omega t)$. 

In order to keep the expressions formally similar to those in \cite{Birula1996} and \cite{Leonhardt2009}, which use the $(\rr,t)$ representation, time derivatives $\partial_t$, which correspond to multiplication by $-i\omega$ for harmonic fields, will not be taken.

To exploit the simple transformation properties of $\mathbf{F}_{\pm}$ and $\mathbf{G}_{\pm}$ under duality, we will use the following two six-component vectors:
\begin{equation}
	\label{eq:RS}
\F=\frac{1}{\sqrt{2}}\begin{bmatrix}Z\D+i\B\\Z\D-i\B\end{bmatrix},\
\G=\frac{1}{\sqrt{2}}\begin{bmatrix}\E+iZ\HH\\\E-iZ\HH\end{bmatrix}.
\end{equation}
Both vectors in (\ref{eq:RS}) keep the positive helicity component on the upper three elements and the negative helicity component on the lower three elements. We are now ready to start the derivations.

Firstly, we examine the duality transformation properties of Maxwell's equations in a medium with general linear constitutive relations $\F=N\G$ and derive the conditions on $N$ for the medium to be dual symmetric. We implicitly assume that $N$ can in general be a function of space and frequency $N(\rr,\omega)$, and also include losses. Then, we focus on the particular case of the linear constitutive relations induced by a general space-time geometry and verify that they meet the duality conditions. 

Using the Maxwell's curl equations we can write:
\begin{equation}
\label{eq:timeev}
i\partial_t \F=i\partial_t N\G=\begin{bmatrix}\nabla\times&0\\0&-\nabla\times\end{bmatrix}\G.
\end{equation}
In the $\G (\F)$ basis, the duality transformation of equations (\ref{eq:gendual}) and (\ref{eq:gendual2}) is simply the 6$\times$6 matrix:
\begin{equation}
\label{eq:D}
D_{\theta}=\begin{bmatrix}\exp(-i\theta)I&0\\0&\exp(i\theta)I\end{bmatrix},
\end{equation}
where $I$ is the $3\times3$ identity matrix. We now use (\ref{eq:D}) to transform (\ref{eq:timeev})
{
\begin{equation}
D_{\theta}i\partial_t N D^{-1}_\theta D_{\theta}\G=D_{\theta}\begin{bmatrix}\nabla\times&0\\0&-\nabla\times\end{bmatrix}D^{-1}_\theta  D_{\theta}\G.
\end{equation}
}
Since $D_{\theta}$ commutes with $i\partial_t$ and with the block diagonal operator containing the curls, we obtain
\begin{equation}
\label{eq:trans}
i\partial_t D_{\theta}N D^{-1}_\theta\G_{\theta}=\begin{bmatrix}\nabla\times&0\\0&-\nabla\times\end{bmatrix}\G_{\theta},
\end{equation}
where $\G_{\theta}=D_{\theta}\G$. The necessary and sufficient invariance condition that keeps the form of equation (\ref{eq:trans}) in the $\G_{\theta}$ variable the same as the form of equation (\ref{eq:timeev}) in the $\G$ variable is $D_{\theta}ND^{-1}_\theta=N$, that is, that $N$ and $D_{\theta}$ must commute. This happens if and only if $N$ is block diagonal in 3$\times$3 blocks. If we use the definitions in (\ref{eq:RS}) to work back what this block diagonal condition means for a more common form of the constitutive relations:
\begin{equation}
\label{eq:cons}
\begin{bmatrix}Z_0\D\\\B\end{bmatrix}=\begin{bmatrix}\matr{\epsilon}(\rr,\omega)&\matr{\chi}(\rr,\omega)\\\matr{\gamma}(\rr,\omega)&\matr{\mu}(\rr,\omega)\end{bmatrix}\begin{bmatrix}\E\\Z_0\HH\end{bmatrix},
\end{equation}
we obtain that a block diagonal $N$ forces ($(\rr,\omega)$ dependence implicit):
\begin{equation}
\label{eq:restric}
\matr{\epsilon}=\matr{\mu},\ \matr{\chi}=-\matr{\gamma}.
\end{equation}
The restrictions in (\ref{eq:restric}) are the necessary and sufficient conditions for duality invariance (helicity preservation) for a general linear inhomogeneous and bianisotropic media. When they are met, the time evolution of $\G$ reads:
\begin{equation}
\label{eq:timeevdual}
i\partial_t \begin{bmatrix}\matr{\epsilon}-i\matr{\chi}&0\\0&\matr{\epsilon}+i\matr{\chi}\end{bmatrix}\G=\begin{bmatrix}\nabla\times&0\\0&-\nabla\times\end{bmatrix}\G.
\end{equation}

Equation (\ref{eq:restric}) generalizes the results obtained in \S 2.2 of Ref. \onlinecite{Birula1996} for isotropic and inhomogeneous media without magneto-electric couplings. Since in general $N$ depends on the spatial position, Eq. (\ref{eq:restric}) also applies across the boundaries of different bianisotropic media, which generalizes the results for helicity preservation across boundaries of different isotropic and homogeneous media without magneto-electric couplings \cite{FerCor2012p}. 

We now turn our attention to the constitutive relations induced by an empty but curved space-time with a metric $\gmunu$. They were derived in \cite{Plebanski1960}. In the Riemann-Silberstein representation, they can be shown to be \cite{Birula1996}:

\begin{equation}
\label{eq:gf}
(F_+)^n=-\frac{1}{g^{00}}\left(\sqrt{-g}g^{nm}-ig_{0k}\varepsilon^{nkm}\right)(G_+)_m
\end{equation}
where $n,m=[1,2,3]$, $\sqrt{-g}$ is the square root of the determinant of $-\gmunu$, the inverse metric $g^{\mu\nu}$ is such that $g^{\mu\nu}g_{\nu k}=\delta^\mu_k=\mathrm{diag}(1,1,1,1)$, and $\varepsilon^{nkm}$ is the totally antisymmetric three dimensional Levi-Civita symbol.

For the other helicity components $(\mathbf{G}_-,\mathbf{F}_-)$, straightforward algebra leads to:
\begin{equation}
\label{eq:gf2}
(F_-)^n=-\frac{1}{g^{00}}\left(\sqrt{-g}g^{nm}+ig_{0k}\varepsilon^{nkm}\right)(G_-)_m
\end{equation}
We can combine (\ref{eq:gf}) and (\ref{eq:gf2}) in a matrix form:
\begin{equation}
\label{eq:GF}
\F=\begin{bmatrix}\matr{A_+}&0\\0&\matr{A_-}\end{bmatrix}\G,
\end{equation}
where $A_\pm^{nm}=(-\sqrt{-g}g^{nm}\mp ig_{0k}\varepsilon^{nkm})/g_{00}$. From our previous discussion, the block diagonal form of (\ref{eq:GF}) implies duality invariance of a curved space-time. Since we have not imposed any restriction on $\gmunu$, duality invariance must be inherent to the structure of any space-time metric. Duality invariance can hence be seen as a necessary condition for any transformation medium. Furthermore, using results from \cite{Leonhardt2006b}, we arrive at some interesting conclusions. In \cite{Leonhardt2006b}, the authors showed that $\matr{\epsilon}$ is related to space-only transformations and $\matr{\chi}$ to transformations which mix space and time components. In equation (\ref{eq:timeevdual}) we note that, since the curl operator has an opposite sign effect on the two helicity components, $\matr{\epsilon}$ has the same effect in the time evolution of both helicity components while $\matr{\chi}$ causes a helicity dependent transformation. Consequently, from the coordinate transformation point of view, space-only transformations are helicity independent while space time mixing has a helicity dependent effect.

We will now discuss how the requirement of duality invariance in transformation electromagnetics affects the design of metamaterials. If all the constituent meta atoms are dual symmetric, their collective response will also be dual symmetric. In other words, if the field scattered by each meta atom preserves the helicity of its exciting field, helicity will be preserved by the entire meta medium. Dual symmetric meta atoms are the only kind of meta atoms needed to implement a general transformation. We will later discuss two realistically implementable classes of dual symmetric meta atoms. 

At this point, the question arises of whether a dual symmetric metamaterial could be built using non-dual symmetric meta atoms. In other words, if by some collective effect, there is a cancellation of the changed helicity components generated by each non dual meta atom. There is evidence that, in general, a medium composed by several copies of the same particle does not preserve helicity unless the particle itself preserves helicity. For example, \cite{FerCor2012c} shows that random mixtures of small particles are dual only when the particles themselves are dual. Also, for arrays of meta atoms \cite{Papakostas2003,Ren2012}, the type of polarisation conversion obtained in both reflection and transmission shows that helicity was not preserved in the interaction. In those experiments, the amount by which the array rotates the plane of linear polarisation depends on the polarisation angle of the incident light. This negates the conservation of the helicity eigenstates (see \cite{FerCor2012c} for an extended discussion). Duality is therefore not a given. The investigation of the question at the beginning of the paragraph is left for the future. In the remaining of this paper we only consider the engineering of the electromagnetic properties of the individual meta atoms as the way to achieve duality of the metamaterial.

We advance to the study of duality symmetry for the meta atoms. We assume that the size of a meta atom allows to model it by a polarisability tensor and derive the restrictions that the tensor must meet in order for the meta atom to be dual symmetric.

We start with a dipolar scatterer at position $\rrp$ with polarisability tensor $M$. The electric ($\ed$) and magnetic ($\md$) dipoles induced by an incident electromagnetic field can be written as the product of the $6\times6$ tensor $M$ times the six component column vector formed by the incident electric $\Erw$ and magnetic $\Hrw$ fields:
\begin{equation}
\label{eq:M}
\begin{split}
\begin{bmatrix}\ed(\omega)\\\md(\omega)\end{bmatrix}=M&\begin{bmatrix}\Erw\\\Hrw\end{bmatrix}=\\
&\begin{bmatrix}\MdE(\omega) & \MdH(\omega)\\\MmE(\omega) & \MmH(\omega)\end{bmatrix}
\begin{bmatrix}\Erw\\\Hrw\end{bmatrix},
\end{split}
\end{equation}
where $M$ is decomposed into its four $3\times3$ blocks, which are labeled using an obvious notation. The strategy we now follow is to impose that the total scattered field due to $\ed(\omega)$ and $\md(\omega)$ preserves the helicity of the incident field. We first obtain the relationship that must hold between $\ed(\omega)$ and $\md(\omega)$ in order for their combined emission to have a well defined helicity. Then, we find the conditions that $M$ must meet so that incident fields with well defined helicity induce dipoles which produce a scattered field with the same well defined helicity. The $\omega$ dependence is from now on implicit. Strictly speaking, the restrictions that we will obtain must be met across the whole bandwidth of operation.

For the first task, we consider the field emitted by an electric dipole $\ed$ and a magnetic dipole $\md$ located at the same point in an infinite homogeneous and isotropic medium with electric and magnetic constants $(\epsilon,\mu)$. We denote by $(\Eed,\Hed)$ the fields produced by the electric dipole $\ed$ and  $(\Emd,\Hmd)$ those produced by the magnetic dipole $\md$. The total fields are the sum of the fields radiated by the two dipoles
\begin{equation}
	\E=\Eed+\Emd,\ \HH=\Hed+\Hmd,
\end{equation}
from where we can obtain the two helicity components:
\begin{equation}
\begin{split}
	\sqrt{2}\mathbf{G}_+&=\left(\Eed+\Emd\right)+iZ\left(\Hed+\Hmd\right),\\
\sqrt{2}\mathbf{G}_-&=\left(\Eed+\Emd\right)-iZ\left(\Hed+\Hmd\right).
\end{split}
\end{equation}
A total field with well defined helicity equal to +1 will have no component of helicity equal to -1, thus 
\begin{equation}
\label{eq:zero}
\sqrt{2}\mathbf{G}_-=\left(\Eed+\Emd\right)-iZ\left(\Hed+\Hmd\right)=0.
\end{equation}
To solve (\ref{eq:zero}), we use the relations in Chap. 9.3 of Jackson \cite{Jackson1998}: A magnetic dipole $\md$ produces electric and magnetic fields $(\Emd,\Hmd)$ which are related to the electric and magnetic fields  $(\Eedbar,\Hedbar)$ produced by an auxiliary electric dipole $\edbar$ in the following way:
\begin{equation}
	\label{eq:ddmm}
\edbar=\frac{\md}{c},\ \Emd=-Z\Hedbar,\ \Hmd=\frac{1}{Z}\Eedbar.
\end{equation}
Note that, for now, $\edbar$ and $\ed$ are not related. Using (\ref{eq:ddmm}) we turn (\ref{eq:zero}) into
\begin{equation}
	\label{eq:plus}
	\Eed-iZ\Hed = i\Eedbar+Z\Hedbar.
\end{equation}
Equation (\ref{eq:plus}) must be met in all points of space. Since the radiated fields depend linearly on the dipole vectors, the solution is
\begin{equation}
	\frac{\md}{c}=\edbar=-i\ed.
\end{equation}
The corresponding steps for a well defined helicity equal to -1 result in $\md/c=\edbar=i\ed$. 

We conclude that:
\begin{equation}
\label{eq:dm}
	\ed=\pm i\frac{\md}{c}
\end{equation}
are the only two cases when an electric and magnetic dipoles at the same point produce a field with well defined helicity, respectively equal to $\pm 1$. Both types of dipole must be present for it. 

We can now advance to the last part of our program and find the conditions on the polarisability tensor $M$ under which the helicity of the incident field is preserved in the scattered field due to the induced dipoles in (\ref{eq:M}). We proceed by changing our representations of the incident fields and the induced dipoles in equation (\ref{eq:M}) in order to separate the two helicity components. For the fields, we will use the vector $\G$ in (\ref{eq:RS}). For the dipoles, in light of (\ref{eq:dm}), the transformation $\mathbf{q}_{\pm}=1/\sqrt{2}\left(\ed \pm i \md/c\right)$, separates the dipolar components that produce fields with well defined helicity. The transformations to obtain $\G$ from $\E$ and $\HH$, and $\mathbf{q}_{\pm}$ from $\ed$ and $\md$ are respectively given by the matrices
\begin{equation}
		T_1=\frac{1}{\sqrt{2}}\begin{bmatrix}I & iZ\\I & -iZ\end{bmatrix},\\
		T_2=\frac{1}{\sqrt{2}}\begin{bmatrix}I & \frac{i}{c}I\\I & -\frac{i}{c}I\end{bmatrix}.
\end{equation}
With the use of these matrices, we transform equation (\ref{eq:M})
\begin{equation}
	\label{eq:T}
		T_2\begin{bmatrix}\ed\\\md\end{bmatrix}=T_2MT_1^{-1}T_1\begin{bmatrix}\mathbf{E}\\\mathbf{H}\end{bmatrix}
\end{equation}
into
\begin{equation}
	\label{eq:pg}
\begin{bmatrix}\mathbf{q}_+\\\mathbf{q}_-\end{bmatrix}=T_2MT_1^{-1}\G.
\end{equation}

In light of (\ref{eq:pg}), the condition for helicity to be preserved is that $T_2MT_1^{-1}$ must be 3$\times$3 block diagonal, which then imposes:
\begin{equation}
\label{eq:dual}
\MdE=\epsilon\MmH,\ \MmE=-\frac{\MdH}{\mu}.
\end{equation}
Note that this condition does not depend on whether the tensors represent a lossless or lossy meta atom. When $(\ref{eq:dual})$ is met, we obtain 
\begin{equation}
	\label{eq:txdipoledual}
	\begin{split}
	&\begin{bmatrix}\mathbf{q}_+\\\mathbf{q}_-\end{bmatrix}=
	\begin{bmatrix}\MdE-i\sqrt{\frac{\epsilon}{\mu}}\MdH & 0\\0 & \MdE+i\sqrt{\frac{\epsilon}{\mu}}\MdH\end{bmatrix}\G.
	\end{split}
\end{equation}
It is clear from the derivations that, a field with well defined helicity incident upon a small scatterer whose polarisability tensor meets (\ref{eq:txdipoledual}) will only induce the dipole of type (\ref{eq:dm}) that corresponds to its helicity. The resulting scattered field radiated by such dipole will preserve the helicity of the incident field. We conclude that, for scatterers described by their polarisability tensors, the relations in (\ref{eq:dual}) are the necessary and sufficient conditions for helicity preservation, or equivalently, duality symmetry. Therefore, and accordingly to our previous arguments, in the context of transformation electromagnetics and metamaterials, equation (\ref{eq:dual}) provides a constraint of the electromagnetic response of the meta atoms that does not sacrifice the generality of the achievable transformation. In \cite{Karilainen2012}, the authors arrive at conditions (\ref{eq:dual}) as one of the necessary conditions for zero backscattering of an electrically small object.

The comparison of equations (\ref{eq:txdipoledual}) and (\ref{eq:timeevdual}) shows that both $\matr{\epsilon}$ and $\MdE$ have the same action on the two helicity states, while $\matr{\chi}$ and $\MdH$ are the ones responsible for helicity dependent transformations. The spatial inversion properties of the inclusion are crucial to establish a priori which inclusions can and which cannot exhibit non-zero cross-polarisabilities $(\MdH,\MmE)$. For example, if we take inclusions that are invariant under a spatial inversion (parity) operation, their cross-polarisabilities can be shown to vanish due to the transformation properties of $\ed,\md,\E$ and $\HH$ under spatial inversion (Table 6.1 in \onlinecite{Jackson1998}). 

\begin{figure}[h]
\subfloat{\includegraphics[scale=1]{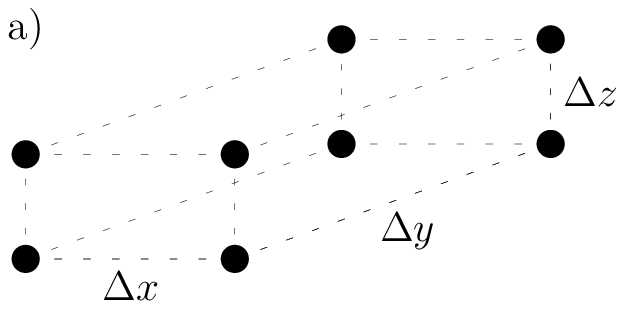}}
\hspace{0.5cm}
\subfloat{\includegraphics[scale=1]{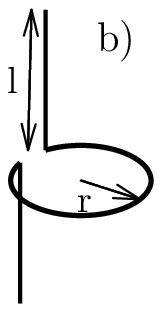}}
\\
\subfloat{\includegraphics[scale=1]{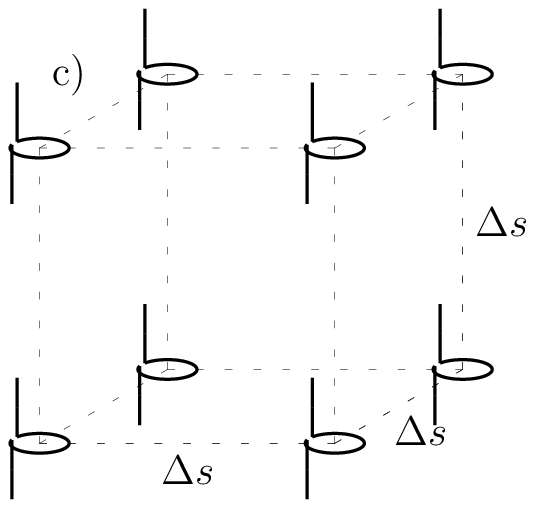}}
\caption{(a,c) Lattice unit cells. (b) Single turn helix. (a) Spheres in an orthorombic lattice arrangement ($\Delta_x\neq \Delta_y\neq \Delta_z$). The spatial inversion symmetry of this structure precludes it from exhibiting magneto electric coupling in its effective constitutive relations. In (c), the inversion symmetry of the cubic lattice is broken by the chiral inclusions (b), and the magneto electric coupling is allowed.}
\label{fig:lattice}
\end{figure}

For the overall effective response of the metamaterial, the properties of the three dimensional arrangement are also important. For example, if we choose a Bravais lattice with sites $\mathbf{r}(n_1,n_2,n_3)$ given by
\begin{equation}
\label{eq:brav}
\mathbf{r}(n_1,n_2,n_3)=n_1\mathbf{a}+n_2\mathbf{b}+n_3\mathbf{c},
\end{equation}
where $n_i$ are integers and $(\mathbf{a},\mathbf{b},\mathbf{c})$ are the lattice vectors, spatial inversion is always a symmetry of the lattice because to each point $(n_1,n_2,n_3)$ there exist its spatially inverted image at $(-n_1,-n_2,-n_3)$. Using now the spatial inversion transformation properties of $\E,\HH,\D$, and $\B$, we can see that the lattice itself cannot induce non-zero values of the constitutive magneto electric tensors $(\matr{\chi},\matr{\gamma})$ in (\ref{eq:restric}). In a Bravais lattice, these tensors must originate from the inclusions cross-polarisabilities $(\MdH,\MmE)$. This situation is analogous to the breaking of time reversal symmetry in a magnetic crystal due not to the lattice itself, but to the alignment of the magnetic moments of the atoms in it and their transformation properties under time reversal. Fig. \ref{fig:lattice} illustrates the discussion about spatial inversion. In this context, and considering Eq. (\ref{eq:txdipoledual}) we can also conclude that $\MdE$ performs space-only transformations and $\MdH$ mixes space and time components. 

We now apply our results to two kinds of inclusions that are commonly considered for metamaterials: Dielectric spheres \cite{Wheeler2010PHDTH,Nieto2011,Soukoulis2011,Liu2012} and conducting chiral inclusions \cite{Semchenko2009,Gansel2009,Soukoulis2011}. Inclusions that are small with respect to the wavelengths of the operating bandwidth can be modeled with good approximation by a pair of colocated electric and magnetic dipoles. In this approximation, the sphere can be made to meet the duality condition (\ref{eq:dual}) by appropriately choosing its radius as a function of the material \cite{Zambrana2013b}. For a conducting helix and other conducting chiral inclusions, duality can be ensured by adjusting their geometrical dimensions. Fig. \ref{fig:lattice} illustrates the two kinds of inclusions.

Spheres have spatial inversion symmetry. Therefore $\MdH=\MmE=0$, and it follows that the sphere is an inclusion that cannot produce helicity dependent transformations for arrangements of the type (\ref{eq:brav}). It also follows that the only condition that a sphere has to meet for it to be dual symmetric is $\MdE=\epsilon\MmH$. Consider then a small dielectric sphere with relative electric and magnetic constants equal to $\epsilon_S$ and 1, respectively. The polarisabilities of a such a sphere when immersed in a homogeneous and isotropic medium can be derived analytically. Their expressions can be found for example in chap. 3.4 of Ref. \onlinecite{Wheeler2010PHDTH}. In our choice of units for the tensors, they read:

\begin{equation}
	\label{eq:mdemmh}
	\MdE=I\epsilon\frac{6\pi i}{k^3}a_1\ \, \ \MmH=I\frac{6\pi i}{k^3}b_1,
\end{equation}
where $\epsilon$ and $k$ are those of the host medium. The numbers $a_1$ and $b_1$ are the Mie coefficients of dipolar order. Expressions for the Mie coefficients can be found for example in Sec. 9.25 of Ref. \onlinecite{Stratton1941}. The duality condition (\ref{eq:dual}) is then met for the sphere when $a_1=b_1$. Assuming that the relative magnetic constants of both the sphere and the host medium are one, the Mie coefficients depend on the wavelength of the illumination and the permittivity $\epsilon_S$ and radius $r_S$ of the sphere. For a given wavelength and $\epsilon_S$, the solution to the equation $a_1(r_S)=b_1(r_S)$ determines one particular radius. For that radius, the sphere is dual in the dipolar approximation according to (\ref{eq:dual}) and (\ref{eq:mdemmh}). Outside the dipolar regime, if higher multipolar orders are considered, the sphere ceases to be dual. There will be some helicity change upon scattering. The idea is that for small spheres, where the non-dipolar terms are very small, the helicity change will be correspondingly small. For example, in a recently published study of duality in dielectric spheres \cite{Zambrana2013b}, the following case can be found: A sphere of 130 nm radius and a refractive index of 2.55 is dual in the dipolar regime ($a_1=b_1$). The total helicity conversion due to symmetry breaking higher multipolar orders is of the order of $10^{-4}$ in converted power.

Now to the other example. Chiral inclusions lack spatial inversion symmetry. Consequently, non-zero electric and magnetic cross-polarisabilities $\MdH$ and $\MmE$ are allowed. This type of inclusions are inherently suitable for the implementation of helicity dependent transformations. Helices and chiral split ring resonators (Ch-SRR) \cite{Marques2007,Marques2007b,Semchenko2009,Gansel2009} are being considered as meta atoms for operation from the microwave to the infra-red regime. Analytical expressions for their polarisability tensors have been derived under suitable approximations, \cite{Tretyakov1996,Marques2007b,Radi2013}. Using those expressions, it can be seen that duality (\ref{eq:dual}) can be achieved around the resonant frequency of the inclusion by adjusting its dimensions. For the helix (Sec. 4.1 of Ref. \onlinecite{Radi2013}), the key dimensional parameter is $r^2/l$, the ratio between the square of the radius of the loop and the length of the straight wire (see Fig. \ref{fig:lattice}-(b)). For the Ch-SRR (Sec. 3 in Ref. \onlinecite{Marques2007b}), it is the ratio between the square of the radius of the rings and the height separation between the two parallel rings composing the chiral inclusion (see Fig 1. in \cite{Marques2007b} for a drawing of a Ch-SRR). In electromagnetic terms, the meaning of the key parameter is very similar in both the helix and the Ch-SRR cases. 

The value of the key parameter that makes the inclusion dual has a $1/\omega$ dependency. If the structure is made dual for the resonant frequency, many physically interesting phenomena occur. For example, in \cite{Semchenko2009}, it is shown that a helix meeting such condition interacts only with one of the circular polarisations, that is, is transparent for the other one. In \cite{Radi2013}, such a helix is shown to maximally interact with given electromagnetic field, extracting the maximum possible power from them. In \cite{Marques2007b} the authors state that, under such condition, a Ch-SRR has several advantages for building negative refractive index metamaterials including wide operation bandwidth and lack of forbidden bands.

All these conditions were found in those works without consideration of the duality symmetry properties of the structure. In our opinion, the fact that all these interesting and apparently useful phenomena occur when the structure is dual is not a coincidence. We think that it is an indication that the consideration of the duality symmetry provides a useful guide for the design of meta atoms for transformation devices.

In this article, we have introduced the theoretical basis and tools for the use of duality symmetry in the design of transformation electromagnetic devices. In particular, we have derived the constrains that the polarisability tensor of a dual symmetric meta atom must meet. Duality symmetry, equivalent to helicity preservation, is already an inherent property of Maxwell's equations in a curved space-time, therefore, the restriction to the duality invariant class of meta atoms does not restrict the implementable space-time transformations. Additionally, we have shown that the space-only part of the coordinate transformation acts equally on both helicity components of the field, while the part that mixes space and time has a helicity dependent effect. Two families of realistically implementable dual symmetric meta atoms have been discussed, one that exhibits electric-magnetic coupling and one that does not. In simple three dimensional periodical arrangements of the meta atoms (Bravais lattices), the helicity dependent effect can only be achieved for meta atoms exhibiting non-zero electric-magnetic cross-polarisabilities. Additionally, we have found that for a pair of colocated electric ($\ed$) and magnetic ($\md$) dipoles to generate a field with well defined helicity equal to $\pm 1$ they must be related as $\ed=\frac{i}{c}\md$ or $\ed=-\frac{i}{c}\md$, respectively. We have also found the restrictions for dual symmetric constitutive relations of inhomogeneous and anisotropic media, which comprise the case of the boundary between two different such media.

{\bf Acknowledgements}
The authors warmly acknowledge Professor Iwo Bialynicki-Birula for very useful discussions during the conception and writing of this paper. This work was funded by the Centre of Excellence for Engineered Quantum Systems (EQuS). G.M.-T is also funded by the Future Fellowship program (FF).


\begin{thebibliography}{32}%
\makeatletter
\providecommand \@ifxundefined [1]{%
 \@ifx{#1\undefined}
}%
\providecommand \@ifnum [1]{%
 \ifnum #1\expandafter \@firstoftwo
 \else \expandafter \@secondoftwo
 \fi
}%
\providecommand \@ifx [1]{%
 \ifx #1\expandafter \@firstoftwo
 \else \expandafter \@secondoftwo
 \fi
}%
\providecommand \natexlab [1]{#1}%
\providecommand \enquote  [1]{``#1''}%
\providecommand \bibnamefont  [1]{#1}%
\providecommand \bibfnamefont [1]{#1}%
\providecommand \citenamefont [1]{#1}%
\providecommand \href@noop [0]{\@secondoftwo}%
\providecommand \href [0]{\begingroup \@sanitize@url \@href}%
\providecommand \@href[1]{\@@startlink{#1}\@@href}%
\providecommand \@@href[1]{\endgroup#1\@@endlink}%
\providecommand \@sanitize@url [0]{\catcode `\\12\catcode `\$12\catcode
  `\&12\catcode `\#12\catcode `\^12\catcode `\_12\catcode `\%12\relax}%
\providecommand \@@startlink[1]{}%
\providecommand \@@endlink[0]{}%
\providecommand \url  [0]{\begingroup\@sanitize@url \@url }%
\providecommand \@url [1]{\endgroup\@href {#1}{\urlprefix }}%
\providecommand \urlprefix  [0]{URL }%
\providecommand \Eprint [0]{\href }%
\providecommand \doibase [0]{http://dx.doi.org/}%
\providecommand \selectlanguage [0]{\@gobble}%
\providecommand \bibinfo  [0]{\@secondoftwo}%
\providecommand \bibfield  [0]{\@secondoftwo}%
\providecommand \translation [1]{[#1]}%
\providecommand \BibitemOpen [0]{}%
\providecommand \bibitemStop [0]{}%
\providecommand \bibitemNoStop [0]{.\EOS\space}%
\providecommand \EOS [0]{\spacefactor3000\relax}%
\providecommand \BibitemShut  [1]{\csname bibitem#1\endcsname}%
\let\auto@bib@innerbib\@empty
\bibitem [{\citenamefont {Leonhardt}(2006)}]{Leonhardt2006}%
  \BibitemOpen
  \bibfield  {author} {\bibinfo {author} {\bibfnamefont {U.}~\bibnamefont
  {Leonhardt}},\ }\href {\doibase 10.1126/science.1126493} {\bibfield
  {journal} {\bibinfo  {journal} {Science}\ }\textbf {\bibinfo {volume}
  {312}},\ \bibinfo {pages} {1777} (\bibinfo {year} {2006})}\BibitemShut
  {NoStop}%
\bibitem [{\citenamefont {Pendry}\ \emph {et~al.}(2006)\citenamefont {Pendry},
  \citenamefont {Schurig},\ and\ \citenamefont {Smith}}]{Pendry2006}%
  \BibitemOpen
  \bibfield  {author} {\bibinfo {author} {\bibfnamefont {J.~B.}\ \bibnamefont
  {Pendry}}, \bibinfo {author} {\bibfnamefont {D.}~\bibnamefont {Schurig}}, \
  and\ \bibinfo {author} {\bibfnamefont {D.~R.}\ \bibnamefont {Smith}},\ }\href
  {\doibase 10.1126/science.1125907} {\bibfield  {journal} {\bibinfo  {journal}
  {Science}\ }\textbf {\bibinfo {volume} {312}},\ \bibinfo {pages} {1780}
  (\bibinfo {year} {2006})}\BibitemShut {NoStop}%
\bibitem [{\citenamefont {Leonhardt}\ and\ \citenamefont
  {Philbin}(2006)}]{Leonhardt2006b}%
  \BibitemOpen
  \bibfield  {author} {\bibinfo {author} {\bibfnamefont {U.}~\bibnamefont
  {Leonhardt}}\ and\ \bibinfo {author} {\bibfnamefont {T.~G.}\ \bibnamefont
  {Philbin}},\ }\href {\doibase 10.1088/1367-2630/8/10/247} {\bibfield
  {journal} {\bibinfo  {journal} {New Journal of Physics}\ }\textbf {\bibinfo
  {volume} {8}},\ \bibinfo {pages} {247} (\bibinfo {year} {2006})}\BibitemShut
  {NoStop}%
\bibitem [{\citenamefont {Plebanski}(1960)}]{Plebanski1960}%
  \BibitemOpen
  \bibfield  {author} {\bibinfo {author} {\bibfnamefont {J.}~\bibnamefont
  {Plebanski}},\ }\href {\doibase 10.1103/PhysRev.118.1396} {\bibfield
  {journal} {\bibinfo  {journal} {Physical Review}\ }\textbf {\bibinfo {volume}
  {118}},\ \bibinfo {pages} {1396} (\bibinfo {year} {1960})}\BibitemShut
  {NoStop}%
\bibitem [{\citenamefont {Leonhardt}\ and\ \citenamefont
  {Philbin}(2009)}]{Leonhardt2009}%
  \BibitemOpen
  \bibfield  {author} {\bibinfo {author} {\bibfnamefont {U.}~\bibnamefont
  {Leonhardt}}\ and\ \bibinfo {author} {\bibfnamefont {T.~G.}\ \bibnamefont
  {Philbin}},\ }\bibfield  {booktitle} {\emph {\bibinfo {booktitle} {Progress
  in Optics}},\ }\href
  {http://www.sciencedirect.com/science/article/pii/S0079663808002023} {\
  \textbf {\bibinfo {volume} {Volume 53}},\ \bibinfo {pages} {69} (\bibinfo
  {year} {2009})}\BibitemShut {NoStop}%
\bibitem [{\citenamefont {Kock}(1956)}]{Kock1956}%
  \BibitemOpen
  \bibfield  {author} {\bibinfo {author} {\bibfnamefont {W.~E.}\ \bibnamefont
  {Kock}},\ }\href {http://www.google.com/patents?id=13lIAAAAEBAJ} {\enquote
  {\bibinfo {title} {Wave refracting devices},}\ } (\bibinfo {year} {1956}),\
  \bibinfo {note} {{U.S. Patent number 2747184}}\BibitemShut {NoStop}%
\bibitem [{\citenamefont {Bialynicki-Birula}(1994)}]{Birula1994}%
  \BibitemOpen
  \bibfield  {author} {\bibinfo {author} {\bibfnamefont {I.}~\bibnamefont
  {Bialynicki-Birula}},\ }\href
  {http://www.cft.edu.pl/\~{}birula/publ/APPPwf.pdf} {\bibfield  {journal}
  {\bibinfo  {journal} {Acta Phys. Pol.}\ }\textbf {\bibinfo {volume} {86}},\
  \bibinfo {pages} {97} (\bibinfo {year} {1994})}\BibitemShut {NoStop}%
\bibitem [{\citenamefont {Bialynicki-Birula}(1996)}]{Birula1996}%
  \BibitemOpen
  \bibfield  {author} {\bibinfo {author} {\bibfnamefont {I.}~\bibnamefont
  {Bialynicki-Birula}},\ }in\ \href
  {http://www.sciencedirect.com/science/article/pii/S0079663808703160} {\emph
  {\bibinfo {booktitle} {Progress in Optics}}},\ Vol.\ \bibinfo {volume}
  {Volume 36},\ \bibinfo {editor} {edited by\ \bibinfo {editor} {\bibfnamefont
  {E.}~\bibnamefont {Wolf}}}\ (\bibinfo  {publisher} {Elsevier},\ \bibinfo
  {year} {1996})\ pp.\ \bibinfo {pages} {245--294}\BibitemShut {NoStop}%
\bibitem [{\citenamefont {Jackson}(1998)}]{Jackson1998}%
  \BibitemOpen
  \bibfield  {author} {\bibinfo {author} {\bibfnamefont {J.~D.}\ \bibnamefont
  {Jackson}},\ }\href@noop {} {\emph {\bibinfo {title} {Classical
  Electrodynamics}}}\ (\bibinfo  {publisher} {Wiley},\ \bibinfo {year}
  {1998})\BibitemShut {NoStop}%
\bibitem [{\citenamefont {Calkin}(1965)}]{Calkin1965}%
  \BibitemOpen
  \bibfield  {author} {\bibinfo {author} {\bibfnamefont {M.~G.}\ \bibnamefont
  {Calkin}},\ }\href {\doibase 10.1119/1.1971089} {\bibfield  {journal}
  {\bibinfo  {journal} {Am. J. Phys.}\ }\textbf {\bibinfo {volume} {33}},\
  \bibinfo {pages} {958} (\bibinfo {year} {1965})}\BibitemShut {NoStop}%
\bibitem [{\citenamefont {Zwanziger}(1968)}]{Zwanziger1968}%
  \BibitemOpen
  \bibfield  {author} {\bibinfo {author} {\bibfnamefont {D.}~\bibnamefont
  {Zwanziger}},\ }\href {\doibase 10.1103/PhysRev.176.1489} {\bibfield
  {journal} {\bibinfo  {journal} {Phys. Rev.}\ }\textbf {\bibinfo {volume}
  {176}},\ \bibinfo {pages} {1489} (\bibinfo {year} {1968})}\BibitemShut
  {NoStop}%
\bibitem [{\citenamefont {Tung}(1985)}]{Tung1985}%
  \BibitemOpen
  \bibfield  {author} {\bibinfo {author} {\bibfnamefont {W.-K.}\ \bibnamefont
  {Tung}},\ }\href@noop {} {\emph {\bibinfo {title} {Group Theory in
  Physics}}}\ (\bibinfo  {publisher} {World Scientific},\ \bibinfo {year}
  {1985})\BibitemShut {NoStop}%
\bibitem [{\citenamefont {Fernandez-Corbaton}\ \emph
  {et~al.}(2013{\natexlab{a}})\citenamefont {Fernandez-Corbaton}, \citenamefont
  {Zambrana-Puyalto}, \citenamefont {Tischler}, \citenamefont {Vidal},
  \citenamefont {Juan},\ and\ \citenamefont {Molina-Terriza}}]{FerCor2012p}%
  \BibitemOpen
  \bibfield  {author} {\bibinfo {author} {\bibfnamefont {I.}~\bibnamefont
  {Fernandez-Corbaton}}, \bibinfo {author} {\bibfnamefont {X.}~\bibnamefont
  {Zambrana-Puyalto}}, \bibinfo {author} {\bibfnamefont {N.}~\bibnamefont
  {Tischler}}, \bibinfo {author} {\bibfnamefont {X.}~\bibnamefont {Vidal}},
  \bibinfo {author} {\bibfnamefont {M.~L.}\ \bibnamefont {Juan}}, \ and\
  \bibinfo {author} {\bibfnamefont {G.}~\bibnamefont {Molina-Terriza}},\ }\href
  {\doibase 10.1103/PhysRevLett.111.060401} {\bibfield  {journal} {\bibinfo
  {journal} {Physical Review Letters}\ }\textbf {\bibinfo {volume} {111}},\
  \bibinfo {pages} {060401} (\bibinfo {year} {2013}{\natexlab{a}})}\BibitemShut
  {NoStop}%
\bibitem [{\citenamefont {Bialynicki-Birula}\ and\ \citenamefont
  {Bialynicka-Birula}(2013)}]{Birula2013}%
  \BibitemOpen
  \bibfield  {author} {\bibinfo {author} {\bibfnamefont {I.}~\bibnamefont
  {Bialynicki-Birula}}\ and\ \bibinfo {author} {\bibfnamefont {Z.}~\bibnamefont
  {Bialynicka-Birula}},\ }\href {\doibase 10.1088/1751-8113/46/5/053001}
  {\bibfield  {journal} {\bibinfo  {journal} {Journal of Physics A:
  Mathematical and Theoretical}\ }\textbf {\bibinfo {volume} {46}},\ \bibinfo
  {pages} {053001} (\bibinfo {year} {2013})}\BibitemShut {NoStop}%
\bibitem [{\citenamefont {Deser}\ and\ \citenamefont
  {Teitelboim}(1976)}]{Deser1976}%
  \BibitemOpen
  \bibfield  {author} {\bibinfo {author} {\bibfnamefont {S.}~\bibnamefont
  {Deser}}\ and\ \bibinfo {author} {\bibfnamefont {C.}~\bibnamefont
  {Teitelboim}},\ }\href {\doibase 10.1103/PhysRevD.13.1592} {\bibfield
  {journal} {\bibinfo  {journal} {Physical Review D}\ }\textbf {\bibinfo
  {volume} {13}},\ \bibinfo {pages} {1592} (\bibinfo {year}
  {1976})}\BibitemShut {NoStop}%
\bibitem [{\citenamefont {Cameron}\ \emph {et~al.}(2012)\citenamefont
  {Cameron}, \citenamefont {Barnett},\ and\ \citenamefont {Yao}}]{Cameron2012}%
  \BibitemOpen
  \bibfield  {author} {\bibinfo {author} {\bibfnamefont {R.~P.}\ \bibnamefont
  {Cameron}}, \bibinfo {author} {\bibfnamefont {S.~M.}\ \bibnamefont
  {Barnett}}, \ and\ \bibinfo {author} {\bibfnamefont {A.~M.}\ \bibnamefont
  {Yao}},\ }\href {\doibase 10.1088/1367-2630/14/5/053050} {\bibfield
  {journal} {\bibinfo  {journal} {New J. Phys.}\ }\textbf {\bibinfo {volume}
  {14}},\ \bibinfo {pages} {053050} (\bibinfo {year} {2012})}\BibitemShut
  {NoStop}%
\bibitem [{\citenamefont {Fernandez-Corbaton}\ \emph
  {et~al.}(2013{\natexlab{b}})\citenamefont {Fernandez-Corbaton}, \citenamefont
  {Vidal}, \citenamefont {Tischler},\ and\ \citenamefont
  {Molina-Terriza}}]{FerCor2012c}%
  \BibitemOpen
  \bibfield  {author} {\bibinfo {author} {\bibfnamefont {I.}~\bibnamefont
  {Fernandez-Corbaton}}, \bibinfo {author} {\bibfnamefont {X.}~\bibnamefont
  {Vidal}}, \bibinfo {author} {\bibfnamefont {N.}~\bibnamefont {Tischler}}, \
  and\ \bibinfo {author} {\bibfnamefont {G.}~\bibnamefont {Molina-Terriza}},\
  }\href {\doibase doi:10.1063/1.4808158} {\bibfield  {journal} {\bibinfo
  {journal} {The Journal of Chemical Physics}\ }\textbf {\bibinfo {volume}
  {138}},\ \bibinfo {pages} {214311} (\bibinfo {year}
  {2013}{\natexlab{b}})}\BibitemShut {NoStop}%
\bibitem [{\citenamefont {Papakostas}\ \emph {et~al.}(2003)\citenamefont
  {Papakostas}, \citenamefont {Potts}, \citenamefont {Bagnall}, \citenamefont
  {Prosvirnin}, \citenamefont {Coles},\ and\ \citenamefont
  {Zheludev}}]{Papakostas2003}%
  \BibitemOpen
  \bibfield  {author} {\bibinfo {author} {\bibfnamefont {A.}~\bibnamefont
  {Papakostas}}, \bibinfo {author} {\bibfnamefont {A.}~\bibnamefont {Potts}},
  \bibinfo {author} {\bibfnamefont {D.~M.}\ \bibnamefont {Bagnall}}, \bibinfo
  {author} {\bibfnamefont {S.~L.}\ \bibnamefont {Prosvirnin}}, \bibinfo
  {author} {\bibfnamefont {H.~J.}\ \bibnamefont {Coles}}, \ and\ \bibinfo
  {author} {\bibfnamefont {N.~I.}\ \bibnamefont {Zheludev}},\ }\href {\doibase
  10.1103/PhysRevLett.90.107404} {\bibfield  {journal} {\bibinfo  {journal}
  {Physical Review Letters}\ }\textbf {\bibinfo {volume} {90}},\ \bibinfo
  {pages} {107404} (\bibinfo {year} {2003})}\BibitemShut {NoStop}%
\bibitem [{\citenamefont {Ren}\ \emph {et~al.}(2012)\citenamefont {Ren},
  \citenamefont {Plum}, \citenamefont {Xu},\ and\ \citenamefont
  {Zheludev}}]{Ren2012}%
  \BibitemOpen
  \bibfield  {author} {\bibinfo {author} {\bibfnamefont {M.}~\bibnamefont
  {Ren}}, \bibinfo {author} {\bibfnamefont {E.}~\bibnamefont {Plum}}, \bibinfo
  {author} {\bibfnamefont {J.}~\bibnamefont {Xu}}, \ and\ \bibinfo {author}
  {\bibfnamefont {N.~I.}\ \bibnamefont {Zheludev}},\ }\href {\doibase
  10.1038/ncomms1805} {\bibfield  {journal} {\bibinfo  {journal} {Nat.
  Commun.}\ }\textbf {\bibinfo {volume} {3}},\ \bibinfo {pages} {833} (\bibinfo
  {year} {2012})}\BibitemShut {NoStop}%
\bibitem [{\citenamefont {Karilainen}\ and\ \citenamefont
  {Tretyakov}(2012)}]{Karilainen2012}%
  \BibitemOpen
  \bibfield  {author} {\bibinfo {author} {\bibfnamefont {A.~O.}\ \bibnamefont
  {Karilainen}}\ and\ \bibinfo {author} {\bibfnamefont {S.~A.}\ \bibnamefont
  {Tretyakov}},\ }\href {\doibase 10.1109/TAP.2012.2207069} {\bibfield
  {journal} {\bibinfo  {journal} {{IEEE} Transactions on Antennas and
  Propagation}\ }\textbf {\bibinfo {volume} {60}},\ \bibinfo {pages} {4449}
  (\bibinfo {year} {2012})}\BibitemShut {NoStop}%
\bibitem [{\citenamefont {Wheeler}(2010)}]{Wheeler2010PHDTH}%
  \BibitemOpen
  \bibfield  {author} {\bibinfo {author} {\bibfnamefont {M.~S.}\ \bibnamefont
  {Wheeler}},\ }\emph {\bibinfo {title} {A scattering-based approach to the
  design, analysis, and experimental verification of magnetic metamaterials
  made from dielectrics}},\ \href
  {https://tspace.library.utoronto.ca/handle/1807/24910} {Ph.D. thesis}
  (\bibinfo {year} {2010})\BibitemShut {NoStop}%
\bibitem [{\citenamefont {Nieto-Vesperinas}\ \emph {et~al.}(2011)\citenamefont
  {Nieto-Vesperinas}, \citenamefont {Gomez-Medina},\ and\ \citenamefont
  {Saenz}}]{Nieto2011}%
  \BibitemOpen
  \bibfield  {author} {\bibinfo {author} {\bibfnamefont {M.}~\bibnamefont
  {Nieto-Vesperinas}}, \bibinfo {author} {\bibfnamefont {R.}~\bibnamefont
  {Gomez-Medina}}, \ and\ \bibinfo {author} {\bibfnamefont {J.~J.}\
  \bibnamefont {Saenz}},\ }\href {\doibase 10.1364/JOSAA.28.000054} {\bibfield
  {journal} {\bibinfo  {journal} {Journal of the Optical Society of America A}\
  }\textbf {\bibinfo {volume} {28}},\ \bibinfo {pages} {54} (\bibinfo {year}
  {2011})}\BibitemShut {NoStop}%
\bibitem [{\citenamefont {Soukoulis}\ and\ \citenamefont
  {Wegener}(2011)}]{Soukoulis2011}%
  \BibitemOpen
  \bibfield  {author} {\bibinfo {author} {\bibfnamefont {C.~M.}\ \bibnamefont
  {Soukoulis}}\ and\ \bibinfo {author} {\bibfnamefont {M.}~\bibnamefont
  {Wegener}},\ }\href {\doibase 10.1038/nphoton.2011.154} {\bibfield  {journal}
  {\bibinfo  {journal} {Nature Photonics}\ }\textbf {\bibinfo {volume} {5}},\
  \bibinfo {pages} {523} (\bibinfo {year} {2011})}\BibitemShut {NoStop}%
\bibitem [{\citenamefont {Liu}\ \emph {et~al.}(2012)\citenamefont {Liu},
  \citenamefont {Miroshnichenko}, \citenamefont {Neshev},\ and\ \citenamefont
  {Kivshar}}]{Liu2012}%
  \BibitemOpen
  \bibfield  {author} {\bibinfo {author} {\bibfnamefont {W.}~\bibnamefont
  {Liu}}, \bibinfo {author} {\bibfnamefont {A.~E.}\ \bibnamefont
  {Miroshnichenko}}, \bibinfo {author} {\bibfnamefont {D.~N.}\ \bibnamefont
  {Neshev}}, \ and\ \bibinfo {author} {\bibfnamefont {Y.~S.}\ \bibnamefont
  {Kivshar}},\ }\href {\doibase 10.1021/nn301398a} {\bibfield  {journal}
  {\bibinfo  {journal} {ACS Nano}\ }\textbf {\bibinfo {volume} {6}},\ \bibinfo
  {pages} {5489} (\bibinfo {year} {2012})},\ \Eprint
  {http://arxiv.org/abs/http://pubs.acs.org/doi/pdf/10.1021/nn301398a}
  {http://pubs.acs.org/doi/pdf/10.1021/nn301398a} \BibitemShut {NoStop}%
\bibitem [{\citenamefont {Semchenko}\ \emph {et~al.}(2009)\citenamefont
  {Semchenko}, \citenamefont {Khakhomov},\ and\ \citenamefont
  {Samofalov}}]{Semchenko2009}%
  \BibitemOpen
  \bibfield  {author} {\bibinfo {author} {\bibfnamefont {I.~V.}\ \bibnamefont
  {Semchenko}}, \bibinfo {author} {\bibfnamefont {S.~A.}\ \bibnamefont
  {Khakhomov}}, \ and\ \bibinfo {author} {\bibfnamefont {A.~L.}\ \bibnamefont
  {Samofalov}},\ }\href {\doibase 10.1007/s11182-009-9249-9} {\bibfield
  {journal} {\bibinfo  {journal} {Russian Phys. J.}\ }\textbf {\bibinfo
  {volume} {52}},\ \bibinfo {pages} {472} (\bibinfo {year} {2009})}\BibitemShut
  {NoStop}%
\bibitem [{\citenamefont {Gansel}\ \emph {et~al.}(2009)\citenamefont {Gansel},
  \citenamefont {Thiel}, \citenamefont {Rill}, \citenamefont {Decker},
  \citenamefont {Bade}, \citenamefont {Saile}, \citenamefont {Freymann},
  \citenamefont {Linden},\ and\ \citenamefont {Wegener}}]{Gansel2009}%
  \BibitemOpen
  \bibfield  {author} {\bibinfo {author} {\bibfnamefont {J.~K.}\ \bibnamefont
  {Gansel}}, \bibinfo {author} {\bibfnamefont {M.}~\bibnamefont {Thiel}},
  \bibinfo {author} {\bibfnamefont {M.~S.}\ \bibnamefont {Rill}}, \bibinfo
  {author} {\bibfnamefont {M.}~\bibnamefont {Decker}}, \bibinfo {author}
  {\bibfnamefont {K.}~\bibnamefont {Bade}}, \bibinfo {author} {\bibfnamefont
  {V.}~\bibnamefont {Saile}}, \bibinfo {author} {\bibfnamefont {G.~v.}\
  \bibnamefont {Freymann}}, \bibinfo {author} {\bibfnamefont {S.}~\bibnamefont
  {Linden}}, \ and\ \bibinfo {author} {\bibfnamefont {M.}~\bibnamefont
  {Wegener}},\ }\href {\doibase 10.1126/science.1177031} {\bibfield  {journal}
  {\bibinfo  {journal} {Science}\ }\textbf {\bibinfo {volume} {325}},\ \bibinfo
  {pages} {1513} (\bibinfo {year} {2009})},\ \bibinfo {note} {{PMID:}
  19696310}\BibitemShut {NoStop}%
\bibitem [{\citenamefont {Zambrana-Puyalto}\ \emph {et~al.}(2013)\citenamefont
  {Zambrana-Puyalto}, \citenamefont {Vidal}, \citenamefont {Juan},\ and\
  \citenamefont {Molina-Terriza}}]{Zambrana2013b}%
  \BibitemOpen
  \bibfield  {author} {\bibinfo {author} {\bibfnamefont {X.}~\bibnamefont
  {Zambrana-Puyalto}}, \bibinfo {author} {\bibfnamefont {X.}~\bibnamefont
  {Vidal}}, \bibinfo {author} {\bibfnamefont {M.~L.}\ \bibnamefont {Juan}}, \
  and\ \bibinfo {author} {\bibfnamefont {G.}~\bibnamefont {Molina-Terriza}},\
  }\href {\doibase 10.1364/OE.21.017520} {\bibfield  {journal} {\bibinfo
  {journal} {Optics Express}\ }\textbf {\bibinfo {volume} {21}},\ \bibinfo
  {pages} {17520} (\bibinfo {year} {2013})}\BibitemShut {NoStop}%
\bibitem [{\citenamefont {Stratton}(1941)}]{Stratton1941}%
  \BibitemOpen
  \bibfield  {author} {\bibinfo {author} {\bibfnamefont {J.~A.}\ \bibnamefont
  {Stratton}},\ }\href@noop {} {\emph {\bibinfo {title} {Electromagnetic
  theory}}}\ (\bibinfo  {publisher} {McGraw-Hill Book Company},\ \bibinfo
  {year} {1941})\BibitemShut {NoStop}%
\bibitem [{\citenamefont {Marqués}\ \emph {et~al.}(2007)\citenamefont
  {Marqués}, \citenamefont {Jelinek},\ and\ \citenamefont
  {Mesa}}]{Marques2007}%
  \BibitemOpen
  \bibfield  {author} {\bibinfo {author} {\bibfnamefont {R.}~\bibnamefont
  {Marqués}}, \bibinfo {author} {\bibfnamefont {L.}~\bibnamefont {Jelinek}}, \
  and\ \bibinfo {author} {\bibfnamefont {F.}~\bibnamefont {Mesa}},\ }\href
  {\doibase 10.1002/mop.22736} {\bibfield  {journal} {\bibinfo  {journal}
  {Microwave and Optical Technology Letters}\ }\textbf {\bibinfo {volume}
  {49}},\ \bibinfo {pages} {2606} (\bibinfo {year} {2007})}\BibitemShut
  {NoStop}%
\bibitem [{\citenamefont {Marques}\ \emph {et~al.}(2007)\citenamefont
  {Marques}, \citenamefont {Mesa}, \citenamefont {Jelinek},\ and\ \citenamefont
  {Baena}}]{Marques2007b}%
  \BibitemOpen
  \bibfield  {author} {\bibinfo {author} {\bibfnamefont {R.}~\bibnamefont
  {Marques}}, \bibinfo {author} {\bibfnamefont {F.}~\bibnamefont {Mesa}},
  \bibinfo {author} {\bibfnamefont {L.}~\bibnamefont {Jelinek}}, \ and\
  \bibinfo {author} {\bibfnamefont {J.~D.}\ \bibnamefont {Baena}},\ }\href
  {http://arxiv.org/abs/0711.4330} {\bibfield  {journal} {\bibinfo  {journal}
  {Proceedings of metamaterial 2007 Conferences, Rome (Italy).
  arXiv:0711.4330}\ ,\ \bibinfo {pages} {214}} (\bibinfo {year}
  {2007})}\BibitemShut {NoStop}%
\bibitem [{\citenamefont {Tretyakov}\ \emph {et~al.}(1996)\citenamefont
  {Tretyakov}, \citenamefont {Mariotte}, \citenamefont {Simovski},
  \citenamefont {Kharina},\ and\ \citenamefont {Heliot}}]{Tretyakov1996}%
  \BibitemOpen
  \bibfield  {author} {\bibinfo {author} {\bibfnamefont {S.}~\bibnamefont
  {Tretyakov}}, \bibinfo {author} {\bibfnamefont {F.}~\bibnamefont {Mariotte}},
  \bibinfo {author} {\bibfnamefont {C.}~\bibnamefont {Simovski}}, \bibinfo
  {author} {\bibfnamefont {T.}~\bibnamefont {Kharina}}, \ and\ \bibinfo
  {author} {\bibfnamefont {J.-P.}\ \bibnamefont {Heliot}},\ }\href {\doibase
  10.1109/8.504309} {\bibfield  {journal} {\bibinfo  {journal} {{IEEE}
  Transactions on Antennas and Propagation}\ }\textbf {\bibinfo {volume}
  {44}},\ \bibinfo {pages} {1006} (\bibinfo {year} {1996})}\BibitemShut
  {NoStop}%
\bibitem [{\citenamefont {Ra'di}\ and\ \citenamefont
  {Tretyakov}(2013)}]{Radi2013}%
  \BibitemOpen
  \bibfield  {author} {\bibinfo {author} {\bibfnamefont {Y.}~\bibnamefont
  {Ra'di}}\ and\ \bibinfo {author} {\bibfnamefont {S.~A.}\ \bibnamefont
  {Tretyakov}},\ }\href {\doibase 10.1088/1367-2630/15/5/053008} {\bibfield
  {journal} {\bibinfo  {journal} {New Journal of Physics}\ }\textbf {\bibinfo
  {volume} {15}},\ \bibinfo {pages} {053008} (\bibinfo {year}
  {2013})}\BibitemShut {NoStop}%
\end{thebibliography}
%
\end{document}